
\magnification\magstep1
\baselineskip=16pt

\def\D{\Delta}

\def\no{\noindent}

\def\D{\Delta}

\def\ga{\gamma}
\def\no{\noindent}

\def\bt{{\bar t}}
\def\cos{{\rm cos}}

\no
\centerline{\bf Phase Transitions of Fermions Coupled to a Gauge Field:}
\centerline{\bf a Quantum Monte Carlo Approach}
\vskip 0.5in
\centerline{M. Hettler$^1$ and K. Ziegler$^2$}
\vskip 0.3in
\no
\centerline{$^1$Department of Physics, University of Florida,
Gainesville, FL 32611, USA}
\centerline{$^2$Institut f\"ur Theorie der Kondensierten Materie,
 Universit\"at Karlsruhe,}
\centerline{ Physikhochhaus, D-76128 Karlsruhe, Germany}
\vskip 0.3in
\noindent
Abstract:\par
\no
A grand canonical system of non-interacting fermions on a square lattice is
considered at zero temperature. Three different phases exist: an empty lattice,
a completely filled lattice and a liquid phase which interpolates between
the other two phases. The Fermi statistics can be changed into a Bose
statistics by coupling a statistical gauge field to the fermions. Using a
quantum Monte Carlo method we investigate the effect of the gauge field on
the critical properties of the lattice fermions. It turns out that there is
no significant change of the phase diagram or the density of particles
due to the gauge field even at the critical points. This result
supports a recent
conjecture by Huang and Wu that certain properties of a three-dimensional flux
line system (which is equivalent to two-dimensional hard-core bosons) can be
explained with non-interacting fermion models.
\vskip6truecm
\vfill
\eject
\no
{\bf 1. Introduction}

\no
Non-interacting fermions play a fundamental role in the theory of exactly
soluble models in one-dimensional quantum systems (e.g. hard-core bosons [1])
and two-dimensional classical systems (e.g. Ising spins [2], dimers [3] and
directed interacting random walks [4]). It was suggested more recently that
non-interacting fermions can also be used to model higher dimensional
systems. Examples are magnets [5] and flux line systems of
the mixed phase in high temperature superconductors with (unphysical)
negative fugacity [6] in $d=3$.
On the other hand, a three-dimensional flux line system with (physical)
positive fugacity is equivalent to a two-dimensional system of hard-core
bosons [7]. Phase transitions (e.g. the Meissner-Abrikosov transition in a
superconductor) exist for positive and for negative fugacities.
Their properties are the same in two-dimensional flux line systems.
This reflects the fact that one-dimensional hard-core bosons are equivalent
to non-interacting fermions. Huang and Wu conjectured that the critical
properties of thermodynamics are also the same [6] in three dimensions. For
instance, the critical exponent for the density of flux lines $n$ at the
transition from the Abrikosov phase ($n>0$) to the Meissner phase ($n=0$)
are the same. It is clear that certain properties of the hard-core bosons
(e.g. superfluidity [8]) are not possible for non-interacting fermions.
Therefore, the equivalence of flux line systems with negative and positive
fugacity cannot be complete in three dimensions.

\no
The connection between fermions and
bosons can be formally understood if we couple a statistical gauge field
to the non-interacting fermions. The gauge field can change the Fermi
statistics to Bose statistics.
A well-known case is the Chern-Simons gauge field which transforms in two
dimensions fermions into effective particles (anyons) [9]. The statistics of
the anyons depends on the coupling constant. Another example for changing the
statistics of fermions into bosons by means of a statistical field coupled to
non-interacting fermions was discussed in Refs. 10-12. In that case two
fermions (e.g., spin-dependent fermions with spin up and down) are bound in
pairs by the gauge field. This construction is not restricted to two
dimensions but can create hard-core bosons from fermions in any dimension.

\no
The effect of the statistical field was studied by means of large $N$ methods
in two extreme regimes: in the dense limit [10,12] and in the dilute limit
[11]. Unfortunately, the properties
in the intermediate regime are not known. In particular, it would be
interesting
to understand whether there is a cross-over behavior from the dilute system to
the dense regime or a transition between two different phases. For this purpose
we performed a Monte Carlo simulation for the fermions in the statistical
field.
The method and some results (density of particles and the phase diagram)
are presented in the following.

The article is organized as follows: The model of free lattice fermions and
the coupling of a statistical gauge field are discussed in Sect. 2. Then we
describe the quantum Monte Carlo method in Sect. 3. The numerical results for
the density of hard-core bosons in $d=2$ (or flux lines in $d=3$) and the phase
diagram are presented in Sect. 4. The results of the Monte Carlo simulation are
compared in Sects. 5, 6 with the results of the free fermions of Sect. 2.
\bigskip
\no
{\bf 2. The Model}

\no
We define the dynamics of fermions on a square lattice with ${\cal N}$ sites by
an inverse lattice propagator $w$ which describes static fermions (resting
particles) $(t,r)\to(t+\D,r)$ and hopping fermions $(t,r)\to(t+\D,r')$. $r$
and $r'$ are the
coordinates on the square lattice and $t$ the (discrete) time.
Subsequently we will also use space-time notation $x=(t,r)$. The matrix
elements of $w$ are
$$w_{t,r;t+\D,r'}=\cases{
1&for $r'=r$\cr
\D J&for $r',r$ nearest neighbors\cr
0&otherwise\cr
}.\eqno (1)$$
The partition sum, which is a generating function for physical
quantities like the particle density, is the fermion determinant
$Z=\det (w+\zeta)$. $\zeta$ is the fugacity for empty sites.
Since we consider temperature $T=0$, the free energy of the
fermions reads as an integral over the Matsubara
frequency $\omega$ and the two-dimensional Brillouin zone $[-\pi,\pi)^2$
$$F=\int_{-\pi}^{\pi}\int_{-\pi}^{\pi}\int_{-\pi}^{\pi}
\log\{e^{i\omega}[1+\bt(\cos k_1+\cos k_2)]+\zeta\}dk_1dk_2d\omega\eqno (2)$$
with $\bt=2\Delta J$.
Critical lines exist for wave vectors $k_1=k_2=0$ (corresponding to a
transition to an empty lattice) and $k_1=k_2=\pi$ (corresponding to a
transition to a completely filled lattice) with $\zeta=1\pm2\bt\ge0$.
The density of fermions on a square lattice reads
$$n=1-\zeta{\partial F\over\partial\zeta}.\eqno(3)$$
According to the definition of $\zeta$ as the fugacity of empty sites the
density is $n=1$ for $\zeta=0$. Equations (2) and (3) lead,
after some straightforward calculations, to
$$n=1-{1\over\pi^2}\int_{-(\zeta+1)/\bt}^{(\zeta-1)/\bt}\Theta(2-|x|)
{\bf K}(1-x^2/4)dx.\eqno (4)$$
$\Theta(2-|x|)$ is the Heaviside step function and ${\bf K}(m)$ the elliptic
integral [13].
The density $n$ vanishes below $\bt=(\zeta-1)/2$ (for $\zeta>1$) and is $n=1$
below $\bt=(1-\zeta)/2$ (for $\zeta<1$).
This is consistent with the critical lines discussed above.
Above the critical lines the density varies continuously between zero and
one. This regime corresponds to a liquid state of fermions. The density of the
fermions is shown
as a function of $\ga=\sqrt{\zeta}$ in Fig.(4).

Usually
a gauge field can be coupled to the lattice fermions by the transformation
$$w_{x,x'}\to u_{x,x'}\equiv e^{i\Phi_{x,x'}}w_{x,x'},\eqno (5)$$
where the phase is anti-symmetric ($\Phi_{x,x'}=-\Phi_{x',x}$) such that
the field $u_{x,x'}$ is Hermitean. The fermions are transformed into hard-core
bosons by choosing a real field $u_{x,x'}$ which is statistically
independent for
each pair of space-time neighboring sites $x$ and $x'$ [10-12].
Although the new field $u$ is not Hermitean we will call
it in this article a gauge field because it couples to the links of the
lattice fermions.

The partition sum of the hard-core bosons can be expressed as
$$Z=\langle{\det}^2 (u + \ga)\rangle_u.\eqno (6)$$
The fermion determinant is squared because the gauge field is coupled to
two independent fermions (e.g. spin $1/2$ fermions). The fugacity is now
$\ga=\sqrt{\zeta}$. The statistical field has zero mean and the variance is
$$\langle u_{t,r;t+\D,r'}^2\rangle_u=w_{t,r;t+\D,r'}.\eqno (7)$$
The partition sum $Z$ in Eq. (6) is the partition sum of flux lines or Bose
world lines (BWLs). This becomes immediately clear if we expand the
determinants and average with respect to the distribution of $u$:
$$\langle\det (u + \ga)^2\rangle_u=\sum_{\pi_1,\pi_2}(-1)^{\pi_1+\pi_2}
\langle\prod_x[u_{x,\pi_1(x)}+\ga\delta_{x,\pi_1(x)}]
[u_{x,\pi_2(x)}+\ga\delta_{x,\pi_2(x)}]\rangle_u$$
$$=\sum_{\pi_1,\pi_2}(-1)^{\pi_1+\pi_2}
\prod_x[w_{x,\pi_1(x)}+\ga^2\delta_{x,\pi_1(x)}]\delta_{\pi_1(x),\pi_2(x)}
=\sum_\pi\prod_x[w_{x,\pi_1(x)}+\zeta\delta_{x,\pi_1(x)}],\eqno (8)$$
where $\pi_j(x)$ are permutations of space-time sites $\{ x\}$.
The last sum is indeed the partition sum of flux lines (or hard-core bosons)
because it takes
into account all possible configuration of paths from $t=0$ to $t=\beta$
which are not intersecting. A flux line (or BWL) element has the weight
$w_{x,x'}$. (For details see Refs. 10-12.)

\no
The density of BWLs at the inverse temperature $\beta$ reads
$$n=1-{\zeta\over Z{\cal N}\beta}{\partial Z\over\partial \zeta}.\eqno (9)$$
The factor $\zeta^{{\cal N}\beta}$ in the partition sum $Z$ can be canceled
by rescaling $Z\to {\bar Z}=\zeta^{-{\cal N}\beta}Z$. Then $\zeta^{-1}$ is the
fugacity of bosons in ${\bar Z}$, and the density reads with the effective
chemical potential $\mu =-\log \zeta$
$$n={1\over {\cal N}\beta}{\partial \log {\bar Z}\over\partial \mu}.
\eqno (10)$$
Eqs. (6), (7) and (10) are the starting points for our Monte Carlo simulation.
\bigskip
\no
{\bf  3. The QMC - Method }

\no
The form (6) of the partition sum allows us to perform a Monte Carlo
- simulation of our system. This can be seen if one chooses as a (discrete)
distribution  $P (u_{x,x'})$  a symmetric $\delta$ - function
$$P (u_{x,x'}) = 1/2 \{\delta (u_{x,x'} + \sqrt{w_{x,x'}}) +
\delta (u_{x,x'} - \sqrt{w_{x,x'}})\}\eqno (11)$$
that satisfies the required properties (7).
Any observable can be calculated by taking the average
$$\langle A \rangle\equiv\langle A\ {\det}^2 (u + \gamma)
\rangle_{u} = {\sum_{\{u_{x,x'} \}}  A (u_{x,x'} , \gamma)\det^{2}
(u + \gamma)\over\sum_{\{u_{x,x'} \}}\det^{2} (u + \gamma)}.\eqno (12)$$
For instance, to obtain the density of bosons $n(\ga,T)$ we
have to set $A=1-\ga Tr[(u+\ga)^{-1}]/2{\cal N}\beta$ according to
Eq. (9).
Of course, one cannot calculate Z as a function of $u_{x,x'}$ and $\gamma$
for a reasonable lattice size. But since there are configurations of the
$u_{x,x'}$ with very different weight with respect to their contribution to the
partition sum, one should search for a way to sum only the main
configurations, i.e., one should sum the weights of the configurations
 with respect to their probability to be realized.

\no
This is done by the Monte Carlo - algorithm. After choosing some initial
configuration $\{u_{x,x'}\}$ we generate a Markov process by flipping
elements $u_{x,x'}$ with respect to their former value with some probability
$W (\{u_{x,x'}\} \rightarrow \{\tilde{u}_{x,x'}\})$. This
probability has to fulfill the property of `detailed balance':
$$W (\{u_{x,x'}\} \rightarrow \{\tilde{u}_{x,x'}\}) p (\{u_{x,x'}\})
= W (\{\tilde{u}_{x,x'}\} \rightarrow \{u_{x,x'}\}) p (\{\tilde{u}_{x,x'}\})
\eqno (13)$$
where $p (\{u_{x,x'}\})$ denotes the equilibrium probability of the
corresponding configuration $\{u_{x,x'}\}$. We used the minimum of some random
number of a rectangular distribution between 0 and 1 and the `acceptance ratio'
R (the squared ratio of the determinants of the new and the old
configurations of $u_{x,x'}$)
 as the transition probability (Metropolis algorithm). The acceptance ratio R
can be calculated via Green's function, the inverse of the matrix $(u
+ \gamma)_{x,x'}$. If flipping of an element $u_{x,x'}$ is accepted
we can update Green's function with low computational effort.
We are repeating this procedure ('sweeping'
through the space-time lattice) until we are sure to be close to the
equilibrium configuration. Then we can begin to measure the desired quantity
and repeat this as long as necessary to produce stable results with low
statistical error.
\bigskip
\no
This QMC - algorithm is related to the usual fermionic
QMC - method established by Blankenbecler et al. [14,15]. Since
the acceptance ratio is a square, it is always positive and the fermionic sign
problem does not appear, equivalent, e.g., to the fermionic Hubbard model
at half filling.
The main problem of this QMC - method lies in the fact that the flipping
variable $u_{x,x'}$ is fixed to the bonds of the lattice and not to the sites.
Obviously, the number of
flipping variables is $(2 d + 1)\cal N $, since a particle can either hop
($2d$) or stay at its spatial point (1), but it has to go on in the time
direction. This restricts the simulations to small lattice sizes (max.
$\cal N$ $= 100$ sites).

\no
The procedure has been checked by comparing the kinetic energy at half
density with the exact solution ($d = 1$) [16], exact diagonalization
and QMC - data [17,18] of
the quantum spin- 1/2 XY- model to which the hard-core bosons can be
mapped [19]. We find good agreement which
establishes the validity of our algorithm.
\bigskip
\no
{\bf 4. Numerical results}
\bigskip
\no
i) Phase transitions at $J$ = 1

\no
The numerical work was mostly done on the Siemens VP-600EX of the
RZ Karlsruhe. A typical job
for the 8$\times$8 lattice at low temperatures needs about one hour CPU time.
The results shown are for  a simple square lattice in dimension $d = 2$.
We also simulated bosons on a triangular lattice. After
adjusting the hopping strength we find that the system behaves as in the case
of a square lattice. This was expected since there is no frustration in both
cases.

\no
First we consider the dependence of the density on the fugacity
$\gamma$
which is the natural parameter for the simulation and from which
we can deduce the dependence of the density on the bosonic chemical potential
$\mu$.
We expect from the results in [10,12] that a superfluid- insulator transition
takes place at the critical
value $\gamma_{c}$ = 1 at $T=0$ in an infinite system. Due to finite
lattice sizes and finite temperature $T$ in our QMC--simulation
the sharp transition is rounded.
To estimate whether a sharp phase transition really exists we must extrapolate
our data to  $T=0$ and $\cal N\rightarrow
\infty$. This is shown in Figs. 1 and 2 .
The curves for $\gamma < \gamma_{c}$ saturate at a finite density below
a $\ga$--dependent temperature (Fig. 1). In contrast, the graph
for $\gamma_{c}$ does not saturate but decreases monotonically with dominant
linear behavior which can be extrapolated to a finite density at $T=0$.
This behavior is characteristic for $\ga_c$ for a given hopping strength
$J$ and can be used to determine the phase boundaries (see below).
The graphs for $\gamma > \gamma_{c}$ (only $\ga =1.005$ is shown)
decrease faster and  the density becomes exponentially small
as $T\rightarrow 0$ even for finite lattice sizes.

\no
To analyze finite size effects
we look at the density for different lattice sizes $\cal N$ (Fig. 2).
As expected, the density generally decreases upon increasing the lattice size.
The density for $\gamma = .995 < \gamma_{c}$  is clearly
finite in the thermodynamic limit ($\cal N\rightarrow\infty$), even at $T = 0$.
However, the remaining density at $T = 0$ for the critical value
$\gamma_{c}$ = 1 scales to zero  as $1/\cal N$ (see inset of Fig. 2)
within numerical accuracy.
\bigskip
\no
We therefore conclude that the numerical data indicate a second order
phase transition at $T = 0$ in the thermodynamic limit for $J$ = 1 at
$\gamma_{c} = 1$, in accord with analytical results [10,12].
\bigskip
\no
ii) Phase diagram

\no
To obtain the phase diagram in the $\ga$ - $J$ - plane we have to vary
$J$ and look for the value of $\gamma$ at which the density
is linearly decreasing at low temperatures but extrapolates to a finite
remaining density at $T = 0$. Since we do not know that value a priori
we have to do
a considerable amount of jobs to estimate accurate data. Due to limited
CPU-time available we restrict ourselves to the 4$\times$4 lattice to calculate
the shape of the phase boundaries and then check some points in the 8$\times$8
lattice. We find no significant changes of the boundaries in the bigger system.
\bigskip
\no
First we observe a monotonic shrinking of the superfluid region as the
hopping $J$ is decreased as can be seen from Fig. 3. At hopping $J$ = 0,
we know that there is a \underbar{first} order phase transition between
the commensurate phases with $n = 0$ and $n = 1$.
For very small hopping ($J<$.02) the superfluid
region becomes too small to distinguish the numerical data (with statistical
error less than 2\%) from the exact data at zero hopping even for very low
temperatures.
A possible phase diagram might show an extension of this first
order point to values of $J>0$, forming a first order transition line
separating the two phases commensurate with the lattice.
The numerical data do not rule out such an extension, but if it exists it
must be very small ($J_{max} < .02$). We assume in the reasoning below that
there is no extension, but the expressions can be easily modified if this
is not valid. Therefore, we obtain the phase diagram shown as Fig. 3.
\no
The Monte Carlo data (circles) fit  well with the phase boundaries
for a system with noninteracting fermions. Especially the transitions
from the empty lattice to a finite density are in very good
agreement with each other, even for much larger values of $J$ than shown.
The transitions from finite density to an insulating state with density $n=1$
are in good agreement for moderate $J$, but deviations become visible
at about $J=1$. Not only the phase boundaries, but also the density profiles
$n$ vs. $\ga$ of hard-core bosons and free fermions are in good agreement
for small
hopping, see Fig. (4). For larger hopping $J$ there are  deviations,
e.g. the curvature of the boson and fermion data is different for
intermediate densities. Still, the qualitative agreement is quite good.

\no
We found a second order transition at finite $J$ but a density jump of
the size of unity for $J = 0$. This suggests
a scaling ansatz for the density
$$n \propto ({\ga  - \ga_{c}(J)\over J })^{ \eta} \eqno (14)$$
with $\eta =1$ to fit the second order transition
just to the right of the first order
 point at $J_c=0$. In the vicinity of $J_{c}$ the behavior (14) should
be dominant. Indeed, we observe a strong increase of the (negative) slope
of the density profile for decreasing $J$ (Fig. 5). If we plot the
slope from the extrapolation of our data to $T = 0$,
we find indeed a linear dependence of the slope on the inverse hopping $1/J$
(see inset of Fig. 5) for moderate values of $J$.
This is in agreement with the ansatz (14).

\bigskip
\no
{\bf 5. Conclusions}

\no
The numerical results give a complete picture of the
phase transitions in a pure grand canonical system of three-dimensional flux
lines (or two-dimensional hard-core bosons) on a lattice. We found no sign
for a phase which is related to a special gauge field configuration
(e.g., a ``flux phase'' found for the $t-J$ model [20]).
The data indicate second order phase transitions at nonzero hopping
$J$ (with the caveat about the first order line for very small hopping)
from an empty system (density $n = 0$)
to a superfluid phase with noninteger $n$ and finally to an
insulating phase with density $n=1$ as we increase the
chemical potential $\mu$ (decrease $\zeta$ or $\ga$).
This behavior was also found for the free fermion system
given in Eq. (4). The $\ga - J$ - phase diagram of the free fermions agrees
with that of the hard-core bosons within numerical errors (Fig. (3)).
No sign for a singular behavior of the boson system between the dense and
the dilute regime was observed.
A reasonably good agreement of fermion and boson systems is also found for the
density profile
(c.f. Fig. 4). Therefore, we conclude that the thermodynamic properties
(i.e., quantities which can be derived from (2) and (6), respectively) are
not affected by the gauge field fluctuations $u$. This supports the conjecture
of Huang and Wu that the critical properties of higher dimensional systems
(here: three-dimensional flux lines and two-dimensional hard-core bosons) can
be described by a free fermion statistics. This relation offers a new approach
for the statistics of flux-lines (or hard-core bosons) in random potential
where we expect new physical states like the vortex glass [21] and the Bose
glass [22,23].
For instance, the freezing of the fermionic line dynamics due to disorder [24]
might be related to the vortex glass transition of physical flux lines.
The analogy of free fermions and flux lines for $d>2$ would also allow the
extension of the two-dimensional results, where the creation of overhangs and
finite loops of flux lines by replica symmetry breaking was found [25].

\vfill
\eject
{\bf References}
\bigskip

[1] E.H. Lieb and D.C. Mattis, \sl Mathematical Physics in One Dimension

{\ \ \ \ }\rm (Academic Press, New York and London, 1966)

[2] T. Schultz, D.C. Mattis and E.H. Lieb, Rev.Mod.Phys. 36, 856 (1964)

[3] J.Nagle, C.S.O. Yokoi and S.M.Bhattacharjee, in {\sl Phase Transitions and
Critical

{\ \ \ } Phenomena}, ed. C.Domb and J.Lebowitz,
Vol.14 (Academic, London, 1991)

[4] P.G. de Gennes, J.Chem.Phys. 48, 2257 (1968);

{\ \ \ \ } V.L. Prokovsky and S.L. Talapov, Phys.Rev.Lett. 42, 65 (1979)

[5] P. Orland, Nucl.Phys B372, 635 (1992)

[6] H.Y. Huang and F.Y. Wu, Physica A 205, 31 (1994)

[7] D.R. Nelson and H.B. Seung, Phys.Rev. B 39, 9153 (1989)

[8] V.N. Popov, \sl Functional Integrals and Statistical Mechanics

{\ \ \ \ }\rm  (Cambridge University Press, New York, 1987)

[9] E. Fradkin, \sl Field Theories of Condensed Matter Systems

{\ \ \ \ }\rm (Addison-Wesley, Redwood City, 1991)

[10] K. Ziegler, Journ.Stat.Phys. 64 (1/2), 277 (1991)

{\ \ \ \ }M.V. Feigel'man and K. Ziegler, Phys.Rev. B46, 6647 (1992)

[11] K. Ziegler, Physica A 208, 177 (1994)

[12] K. Ziegler, Europh.Lett. 9, 277 (1989)

[13] M. Abramowitz and I.A. Stegun, {\sl Handbook of Mathematical Functions}

{\ \ \ \ } (Dover, New York) (1965)

[14] R. Blankenbecler, D.J. Scalapino and R.L. Sugar, Phys.Rev. D34, 2278
(1981)

[15] E.Y. Loh Jr. and J.E. Gubernatis, \sl Stable numerical simulations

{\ \ \ \ }of models of interacting electrons in Condensed-Matter Physics\rm\
in

{\ \ \ \ }\sl `Electrons Phase Transitions'\rm\  ed. by W. Hanke and Y.V.
Kopaev

{\ \ \ \ }North Holland Physics (Elsevier, New York, 1990)

[16] S. Katsura, Phys. Rev. 127, 1508 (1962)

[17] Y. Okabe and M. Kikuchi, J. Phys. Soc. Jpn. 57, 4351 (1988)

[18] Y. Okabe and M. Kikuchi, J. Phys. Soc. Jpn. 58, 679 (1989)

[19] T. Matsubara and H. Matsuda, Prog. Theor. Phys. 16, 416 (1955),

{\ \ \ \ } Prog. Theor. Phys. 16, 569 (1955)

[20] I. Affleck and J.B. Marston, Phys. Rev. B 37, 3774 (1988)

[21] G. Blatter, M.V. Feigel'man, V.B. Geshkenbein, A.I. Larkin and V.M.
Vinokur,

{\ \ \ \ }(to be published in Rev.Mod.Phys.)

[22] M.P.A. Fisher, D.S. Fisher, G. Grinstein and P.B. Weichman,

{\ \ \ \ }Phys. Rev. B40, 546 (1989)

[23] L. Zhang and M. Ma, Phys. Rev. B45, 4855 (1992)

[24] K. Ziegler, Z.Phys. B84, 17 (1991)

[25] K. Ziegler and A.M.M. Pruisken, Phys. Rev. E51, (1995)

\vfill
\eject
\no
{\bf Figure Captions}

\no
Fig. 1: Density of bosons versus temperature for various fugacities
$\ga$ on a ${\cal N}=8\times 8=64$ lattice
(hopping $J=1$). The density saturates for $\ga < \ga_c = 1$ and becomes
exponentially small for $\ga > \ga_c$ as temperature $T\rightarrow 0$.
For $\ga = \ga_c$ the density drops linearly and approaches a finite, lattice
size
dependent value. This behavior is characteristic and allows us to find the
critical $\ga_c$ for arbitrary hopping strength $J$. $T_o$ is a reference
temperature.
\smallskip

\no
Fig. 2: Density of bosons versus temperature for various lattice sizes $\cal N$
for the fugacity $\ga = 0.995$ and $\ga = 1$,
respectively (hopping strength $J=1$). The
density is a decreasing function of lattice size for all temperatures.
Extrapolating
to $T=0$ gives density values which stay finite as $\cal N$ is growing
for $\ga < \ga_c$. For $\ga_c =1$ the density at $T=0$ behaves like $1/\cal N$
and extrapolates to zero as $\cal N\rightarrow\infty$ (see inset). This
supports the picture of a second order  phase transition at $T=0$ and
$\ga_c=1$.
$T_o$ is a reference temperature.
\smallskip

\no
Fig. 3: Phase diagram  of hard-core bosons and free fermions  on a square
lattice in the $\ga - J $ plane. The circles represent the QMC data for
hard-core bosons, the dashed lines are the phase boundaries for noninteracting,
spinless fermions. The
phases are insulating (Mott insulator) above the upper
(density $n\equiv 0$) and below the lower phase boundary ($n\equiv 1$). In
between,
the particles are in a (super-) fluid phase. The agreement of the phase
boundaries of the two systems is good for small hoppings $J$
 and also for large $J$ for the upper boundaries. The lower boundaries
deviate for larger $J$.
\smallskip

\no
Fig. 4: Density profile for various hoppings $J$ for hard-core bosons
and free fermions on a square lattice. One clearly observes the (insulating)
phases with density $n\equiv0$ and $n\equiv1$, respectively. There is
qualitative
agreement of bosons (symbols) and fermions (solid lines) for small hoppings
but obvious deviations  for larger hopping $J$. The temperatures are in units
of the reference temperature $T_o$.
\smallskip

\no
Fig. 5: Density versus fugacity $\ga$ for three intermediate values
of hopping $J$.
The slope $- dn/d\ga$ increases upon decreasing $J$. The inset shows that
the slope is roughly proportional to the inverse hopping $1/J$.
The value of the slopes in the inset are determined  from the extrapolation
of the data to $T=0$. The dashed lines are a guide to the eye. The temperature
is $T = 1/6 T_o$.
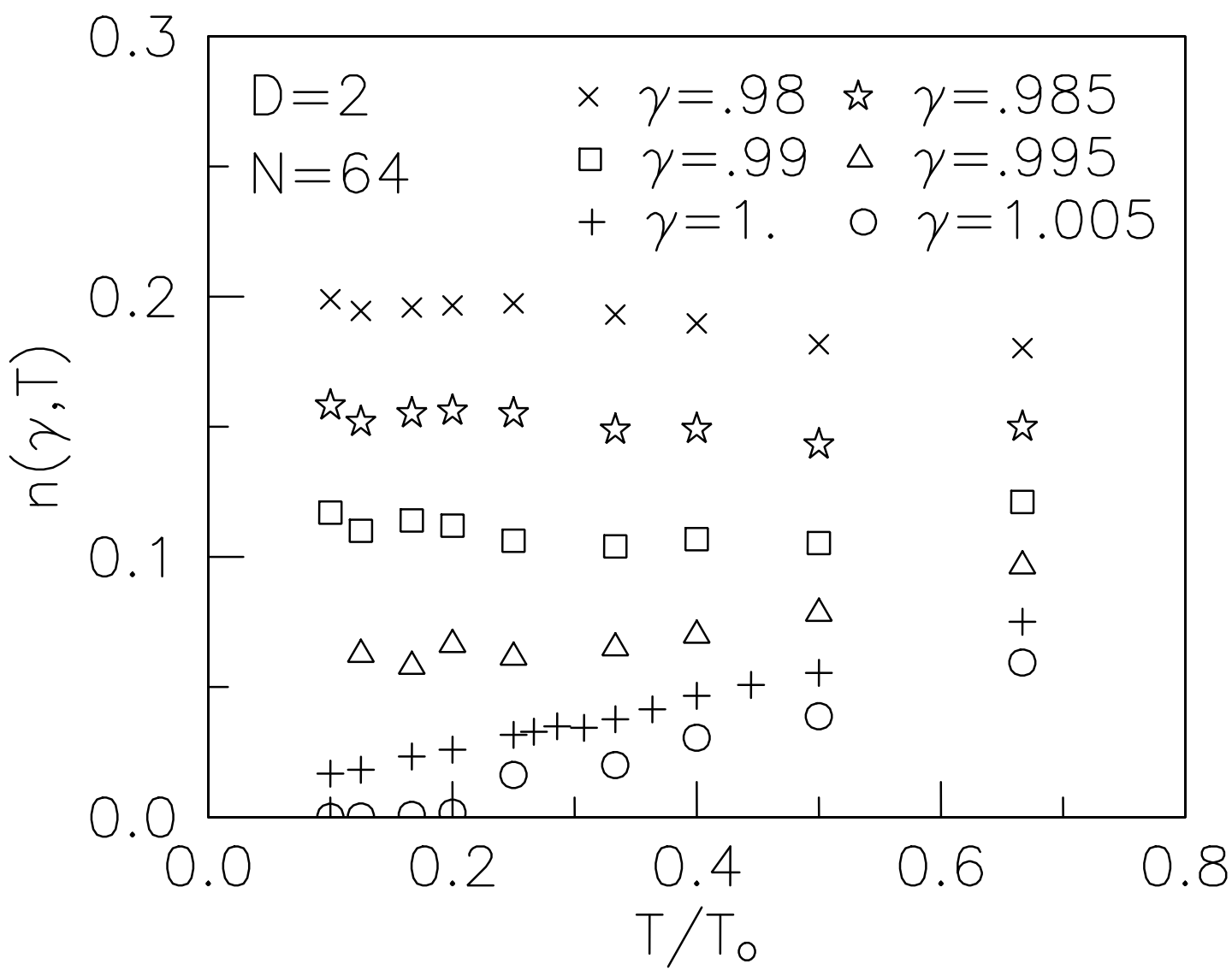
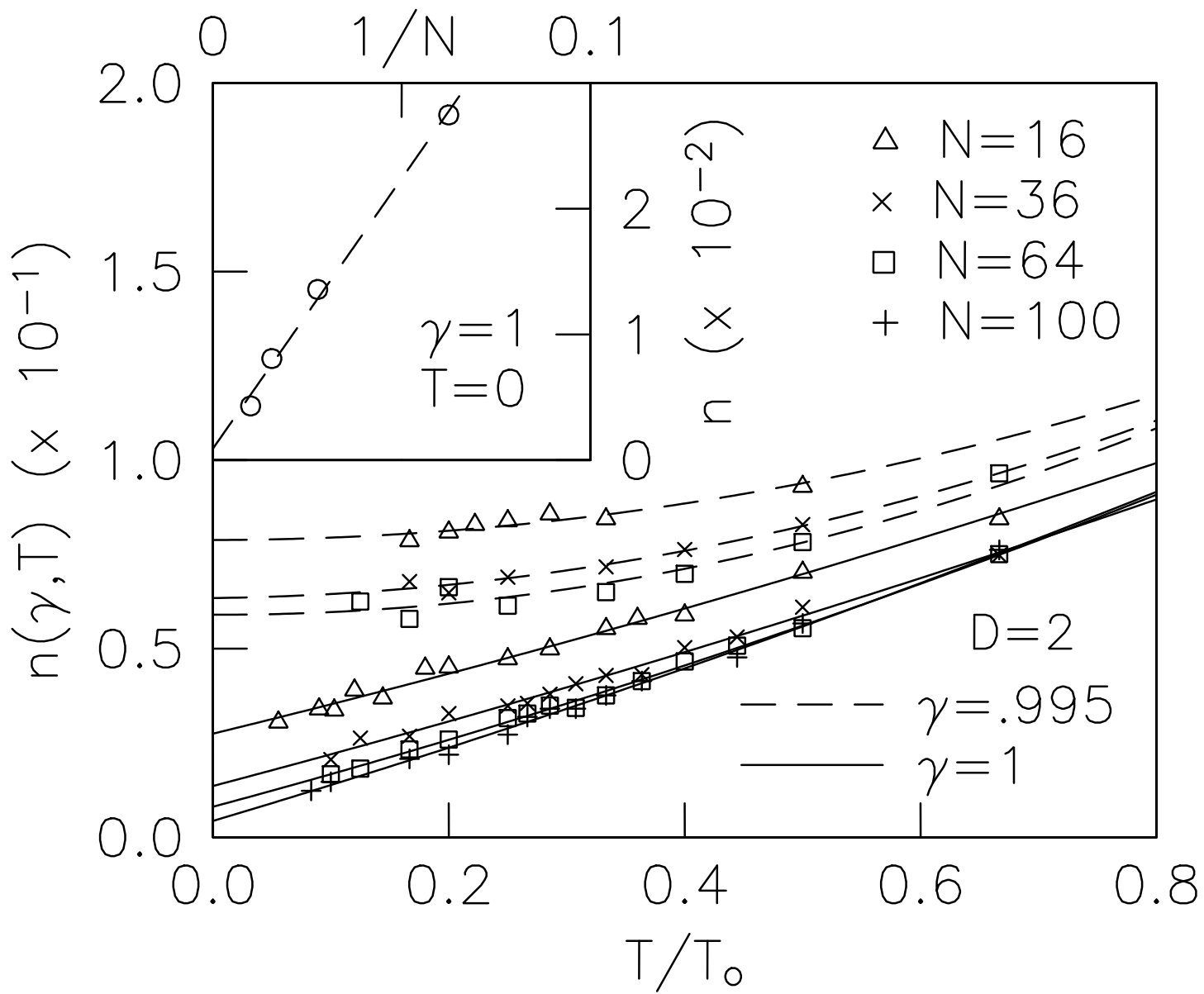
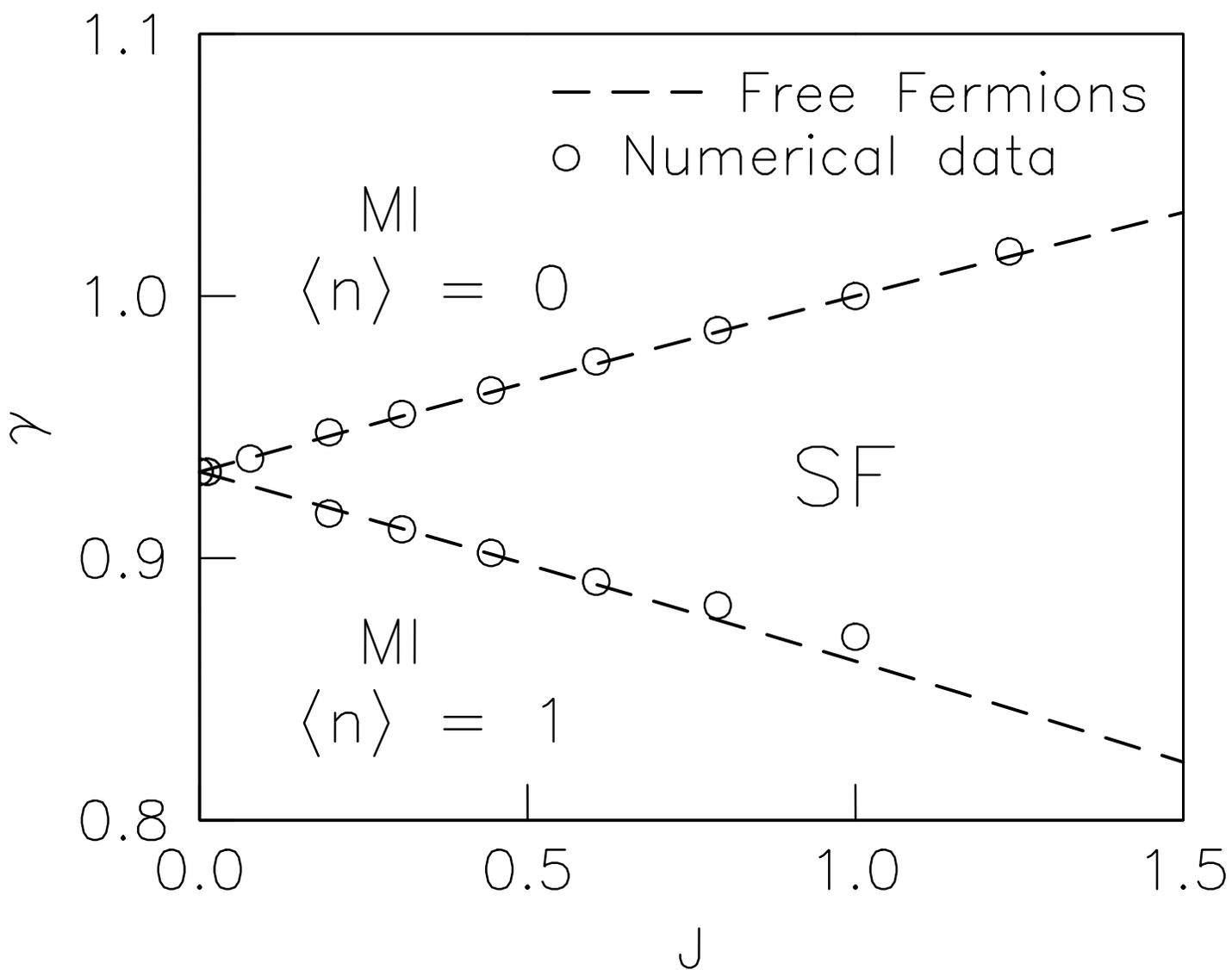
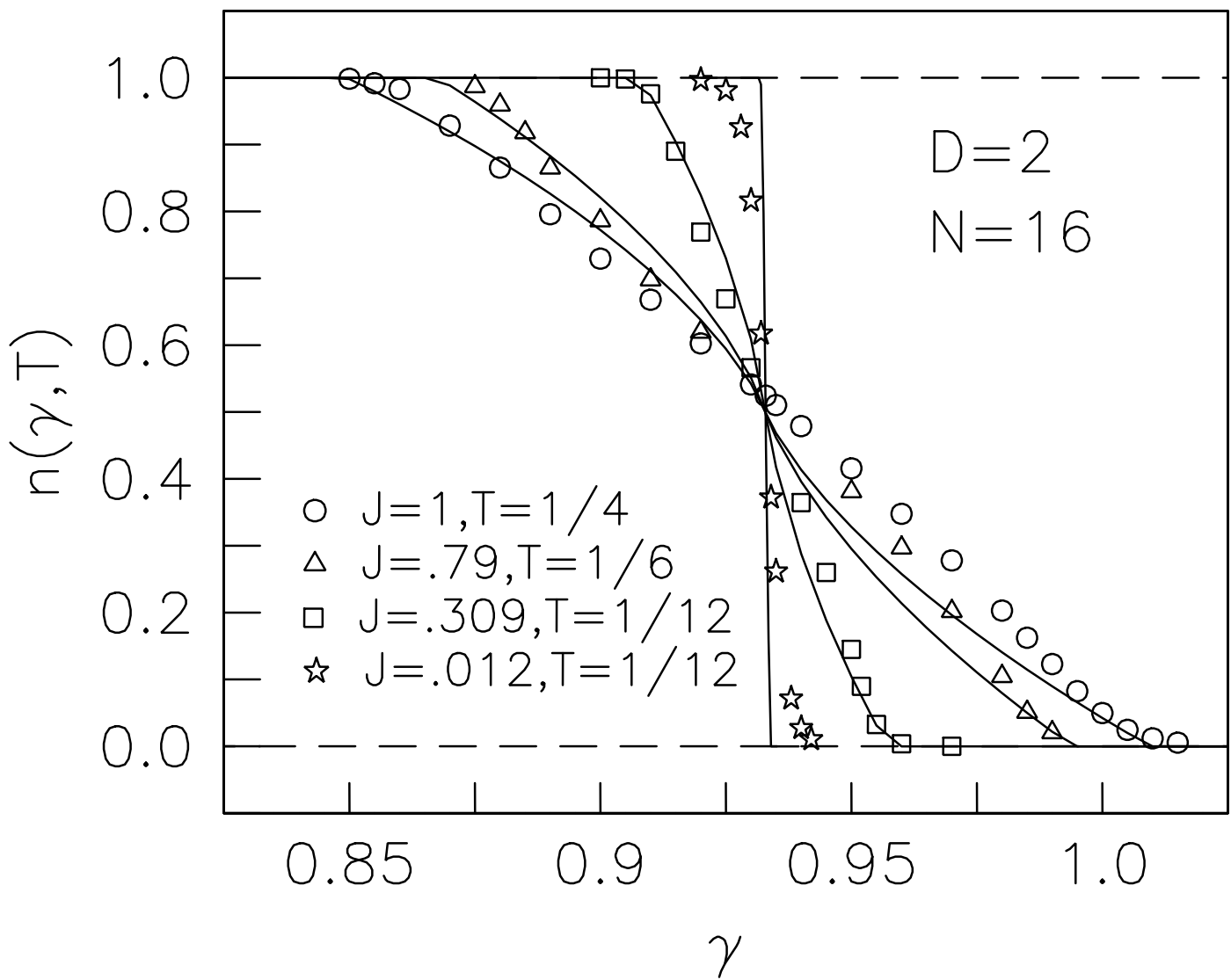
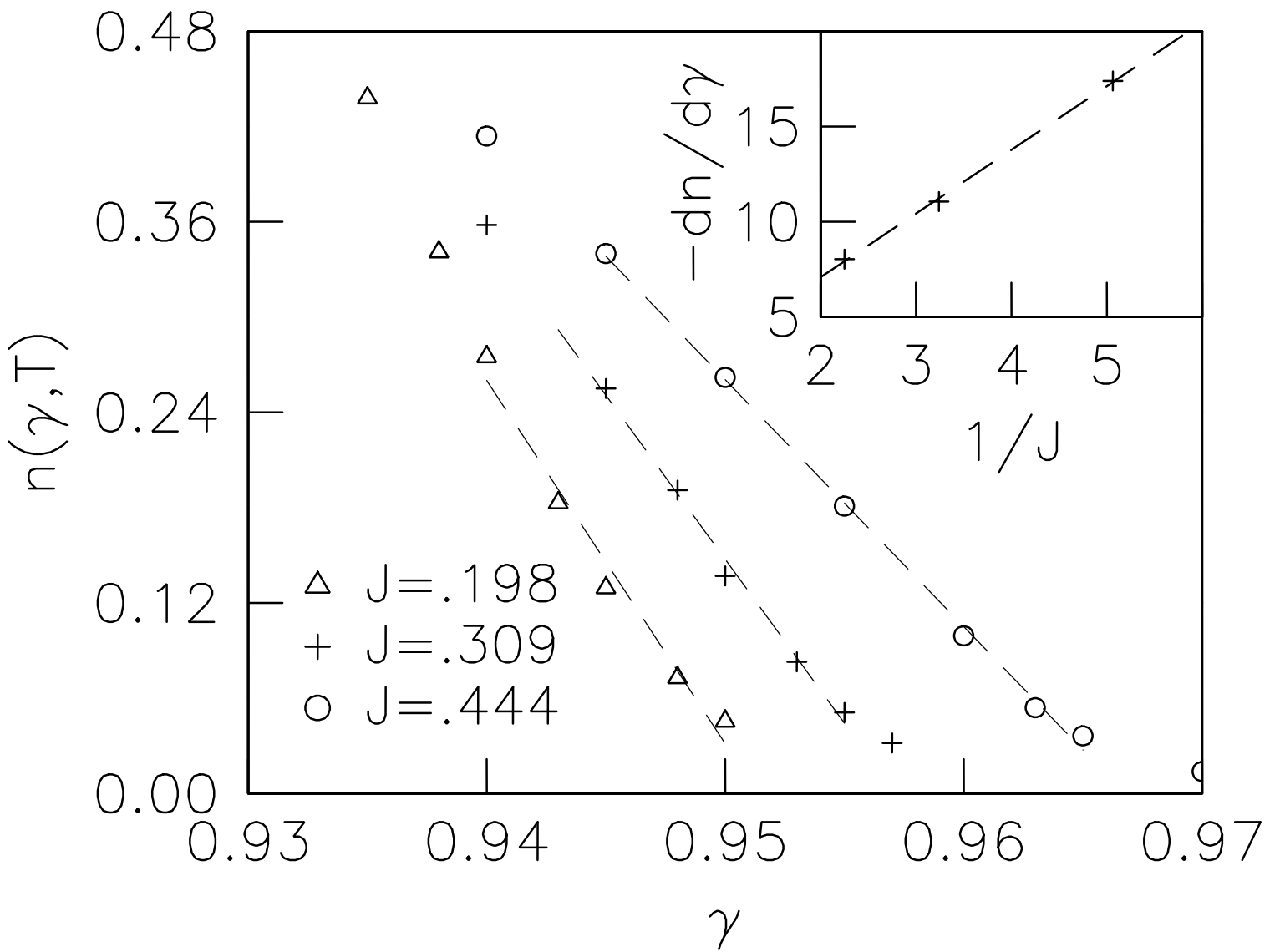
\bye